\documentclass{article}
\usepackage[T1]{fontenc}
\usepackage[utf8]{inputenc}
\usepackage{prettyref}
\usepackage{amstext}
\usepackage{graphicx}
\usepackage[authoryear]{natbib}

\makeatletter
\usepackage{prettyref}
\newrefformat{eq}{eqn\nolinebreak\,\nolinebreak(\ref{#1})}
\newrefformat{sec}{Section\nolinebreak\,\nolinebreak\ref{#1}}
\newrefformat{cha}{Chapter\nolinebreak\,\nolinebreak\ref{#1}}
\newrefformat{tab}{Table\nolinebreak\,\nolinebreak\ref{#1}}
\newrefformat{fig}{Figure\nolinebreak\,\nolinebreak\ref{#1}}
\newrefformat{lem}{Lemma\nolinebreak\,\nolinebreak\ref{#1}}
\newrefformat{thm}{Theorem\nolinebreak\,\nolinebreak\ref{#1}}
\newrefformat{axm}{Axiom\nolinebreak\,\nolinebreak\ref{#1}}
\newrefformat{app}{Appendix\nolinebreak\,\nolinebreak\ref{#1}}
\newrefformat{exe}{Exercise\nolinebreak\,\nolinebreak\ref{#1}}
\newrefformat{cor}{Corollary\nolinebreak\,\nolinebreak\ref{#1}}
\newrefformat{pro}{Property\nolinebreak\,\nolinebreak\ref{#1}}
\newrefformat{def}{Definition\nolinebreak\,\nolinebreak\ref{#1}}
\newrefformat{subsec}{Section\nolinebreak\,\nolinebreak\ref{#1}}
\usepackage{amsmath}
\usepackage{bm}
\numberwithin{equation}{section}
\numberwithin{figure}{section}
\numberwithin{table}{section}
\newcommand{\br}[1]{\mathbf{#1}} 
\newcommand{\bb}[1]{\mathbb{#1}} 
\newcommand{\ca}[1]{\mathcal{#1}} 
\newcommand{\op}[1]{\ca{#1}} 
\usepackage{pifont}
\usepackage{txfonts}
\usepackage{charter}
\date{}

\makeatother

\begin{document}

\title{Leibniz Equivalence, Newton Equivalence, and Substantivalism}

\author{Oliver Davis Johns\thanks{San Francisco State University, Department of Physics and Astronomy,
Thornton Hall, 1600 Holloway Avenue, San Francisco, CA 94132 USA,
<ojohns@metacosmos.org>}}
\maketitle
\begin{abstract}
Active diffeomorphisms map a differentiable manifold to itself. They
transform manifold points and objects without changing the system
of local coordinates used to represent those objects. What has been
called \emph{Leibniz Equivalence} is the assertion that, although
active diffeomorphisms do change manifold objects, they do not change
what is called the \textquotedbl{}physical situation\textquotedbl{}
being modeled by those objects. This paper introduces the contrasting
idea of \emph{Newton Equivalence,} which asserts that the different
values of manifold objects produced by active diffeomorphisms do model
different physical situations. But due to the assumption of general
covariance, these different physical situations are all equally possible.
They represent physically different situations all of which could
happen. This paper compares these two interpretations of active diffeomorphisms,
and comments on their importance in the substantivalism debate. 
\end{abstract}
\rule{0.4\columnwidth}{0.2pt}

$\:$
\begin{enumerate}
\item \emph{Introduction}
\item \emph{Passive Diffeomorphisms}
\item \emph{Active Diffeomorphisms}
\item \emph{Uncontested Points}
\item \emph{Two Examples of Active Diffeomorphisms}
\item \emph{Newton Equivalence}
\item \emph{Leibniz Equivalence}
\item \emph{Substantivalism}
\item \emph{Conclusion}
\end{enumerate}
$\:$

\newpage{}

\section{Introduction}

\label{sec:intro}As he was working toward the field equation of general
relativity, Einstein devised a thought experiment which convinced
him that a generally covariant field equation could have multiple
solutions.\footnote{See Chapter 5 of \citet{torretti} and Chapter 5 of \citet{stachel-btoz}.}
He imagined a special but plausible case in which the energy-momentum
tensor source term would vanish in a region he called a \emph{hole.}
A transformation that was an identity everywhere except the hole would
then modify the solution in the hole region without any change to
the sources. Stachel noted that Einstein had at first referred to
this transformation as what translates as a \emph{coordinate} transformation,
and later, perhaps in response to criticism, as a \emph{point }transformation\emph{.}\footnote{\citet{torretti}, Section 5.6, page 164; \citet{Stachel-active}.}
Stachel suggested that the former term referred to diffeomorphic changes
of local coordinates, and the latter term referred to what he called
active diffeomorphisms, those that transform manifold points without
changing the local coordinate system.\footnote{Using Einstein's form of his thought experiment, \citet{Johns-validity}
outlines a mathematical proof that active diffeomorphisms can produce
multiple solutions of the field equation of general relativity with
the same energy-momentum source. However, it is also suggested that
some of these solutions can be rejected as spurious, leading in some
cases to a unique result.}

In spite of Einstein's hole argument, the final form of his field
equation is generally covariant. Einstein resolved this dilemma by
asserting that the multiple solutions must all represent the same
physical reality. 

A decades old but still influential paper by Earman and Norton\footnote{\citet{earman-norton}. Examples of its influence include recent papers
by \citet{muller} completing, and by \citet{schulman-homotopy} and
\citet{weatherall} refuting, the \textquotedbl{}hole argument.\textquotedbl{}
These papers refer to the \citet{earman-norton} generalization rather
than to Einstein's version of it. See also the current encyclopedia
article, \citet{norton-encyc}.} suggested that Einstein's hole argument should be generalized. The
Earman-Norton generalization makes no use of the detail of Einstein's
special energy-momentum tensor source. Instead, it elevates Einstein's
\emph{resolution} of his dilemma, his assertion that multiple solutions
represent the same physical reality, to a general principle referred
to as \emph{Leibniz Equivalence:} \textquotedbl{}Diffeomorphic models
represent the same physical situation.\textquotedbl{}\footnote{Earman and Norton, \emph{op. cit.,} page 522.}
It is clear from context that the term \textquotedbl{}diffeomorphic
models\textquotedbl{} here refers to models related to each other
by Stachel's active diffeomorphisms.

Whereas Einstein did not address the case of differential equations
other than his own field equation, the Earman-Norton paper asserts
that the principle of Leibniz Equivalence also applies to \textquotedbl{}...Newtonian
spacetime theories with all, one, or none of gravitation and electrodynamics;
and special and general relativity, with and without electrodynamics.\textquotedbl{}\footnote{Earman and Norton, \emph{op. cit.,} page 516.}
Einstein's resolution of his hole argument dilemma is thus generalized
into an assertion about the effect of active diffeomorphisms in essentially
all differential geometric physical models.

The task of the present paper is to suggest that Leibniz Equivalence
is not the only possible interpretation of active diffeomorphisms.
We suggest an alternate interpretation called \emph{Newton Equivalence
}that is more in keeping with current practice in theoretical physics.
In this interpretation, the changes of manifold objects produced by
an active diffeomorphism do change the physical situation. But, due
to general covariance, the new physical situation produced by an active
diffeomorphism, while different, is equally possible. 

Since the dispute here is between two alternate interpretations of
active diffeomorphisms, it is essential to state clearly what the
term means. To establish notation and avoid misunderstandings, Sections
\ref{sec:passive} and \ref{sec:active} review the underlying differential
geometric definitions. \prettyref{sec:summary} then summarizes the
points of agreement between the Leibniz and Newton interpretations,
the basic facts that are to be differently interpreted. 

\prettyref{sec:examples} provides two illustrative examples of the
use of active diffeomorphisms in physics. Sections \ref{sec:newt}
and \ref{sec:leibniz} use those examples to argue for and against
the Newton and Leibniz interpretations. 

\prettyref{sec:substantivalism} discusses Earman and Norton's argument
that substantivalism fails because it requires denial of Leibniz Equivalence.
It is shown that Newton Equivalence, which does deny Leibniz Equivalence,
escapes the unfortunate consequences proposed by these authors. This
section also shows that the attempt to derive indeterminism from the
denial of Leibniz Equivalence fails to generalize the Einstein hole
argument.

\prettyref{sec:conclusion} summarizes the paper's conclusion that
Newton Equivalence is a straightforward and correct interpretation
of active diffeomorphisms, and is consistent with current practices
in theoretical and experimental physics. Also, while choice of the
Leibniz interpretation binds one to a rejection of substantivalism,
choice of the Newton interpretation frees one from that binding and
allows one to remain agnostic on this important issue.\footnote{With the current situation in theoretical physics (for example, the
unknown nature of dark energy, the lack of a satisfactory quantum
theory of gravity), it seems important not to restrict the models
that theorists may try.}

\section{Passive Diffeomorphisms}

\label{sec:passive}Modern differential geometry makes a  distinction
between \emph{manifold objects} and the \emph{coordinate objects}
that represent them in various systems of   local coordinates.\footnote{To make this important distinction clear, throughout this paper manifold
points and objects will be written in \textbf{bold roman type}, while
coordinate objects will use \emph{non-bold italic.}} Thus, given a differentiable manifold $\ca M$ of dimension $m$,
a point $\br x\in\ca M$ is a manifold object represented in different
systems of   local coordinates (here denoted as unprimed or primed
local coordinates)  by
\begin{align}
x & =(x^{1},\ldots,x^{m})=\psi(\br x)\nonumber \\
x' & =(x'^{1},\ldots,x'^{m})=\psi'(\br x)\label{eq:p1}
\end{align}
where $\psi$ and $\psi'$ are different homeomorphic mappings from
$\op M$ to $\bb R^{m}$. The definition of a smooth differentiable
manifold is that both the relation \prettyref{eq:p1a} between any
two systems of   local coordinates, 
\begin{equation}
x'=\psi'\circ\psi^{-1}(x)\label{eq:p1a}
\end{equation}
and its inverse, must be continuously differentiable to arbitrary
order.\footnote{The transformation from $x$ to $x'$ is a diffeomorphic change of
  local coordinates. We will refer to this transformation as a \emph{passive
diffeomorphism, }since it does not change the manifold object $\br x$.
The term \emph{gauge transformation} is also sometimes used in the
literature. This paper takes \textquotedbl{}passive diffeomorphism,\textquotedbl{}
\textquotedbl{}gauge transformation,\textquotedbl{} and \textquotedbl{}diffeomorphic
change of   local coordinates\textquotedbl{} to refer to the same
transformation, \prettyref{eq:p1a}.} 

Functions $\:\br f:\ca M\rightarrow\bb R$ mapping manifold points
$\br x$ to real numbers are manifold objects that have local coordinate
representations $f=\br f\circ\psi^{-1}$ and $f'=\br f\circ\psi'^{-1}$
related by
\begin{equation}
f(x)=\br f(\br x)=f'(x')\label{eq:p2}
\end{equation}
Smooth functions are defined as those for which $f(x)$ in some arbitrary
system of   local coordinates (and therefore in all such systems)
is continuously differentiable to arbitrary order.

Tangent vector fields $\br V(\br x)$ are manifold objects that map
smooth functions to real numbers. These manifold objects are represented
in unprimed and primed   local coordinates, respectively, by 
\begin{equation}
V(x)=\sum_{i=1}^{m}V^{i}(x)\,\partial/\partial x^{i}\;\;\text{and}\;\;V'(x')=\sum_{i=1}^{m}V'^{i}(x')\,\partial/\partial x'^{i}\label{eq:p2a}
\end{equation}
The functions $V^{i}(x)$ and $V'^{i}(x')$ are called the components
of $\br V(\br x)$ in the two systems. The relation between manifold
and coordinate objects is defined as
\begin{multline}
\sum_{i=1}^{m}V^{i}(x)\dfrac{\partial f(x)}{\partial x^{i}}=V(x)f(x)=\br V(\br x)\,\br f(\br x)\\
=V'(x')f'(x')=\sum_{i=1}^{m}V'^{i}(x')\dfrac{\partial f'(x')}{\partial x'^{i}}\label{eq:p3}
\end{multline}
Note that, by nearly universal custom, the operation of tangent vectors
on functions is written in operator form with no parentheses around
the function. The components of a vector field in the unprimed and
primed systems are related by
\begin{equation}
V'^{i}(x')=\sum_{j=1}^{m}\dfrac{\partial x'^{i}}{\partial x^{j}}V^{j}(x)\label{eq:p4}
\end{equation}

Second rank covariant tensor fields are manifold objects $\br g(\br x)$
that bilinearly map an ordered pair of tangent vector fields $\br U(\br x)$
and $\br V(\br x)$ to real numbers, denoted $\br g(\br x)\bigl\{\br U(\br x),\br V(\br x)\bigr\}$.
These manifold objects also have local coordinate representations
\begin{multline}
\sum_{i=1}^{m}\sum_{j=1}^{m}g_{ij}(x)U^{i}(x)V^{j}(x)=\br g(\br x)\bigl\{\br U(\br x),\br V(\br x)\bigr\}\\
=\sum_{i=1}^{m}\sum_{j=1}^{m}g'_{ij}(x')U'^{i}(x')V'^{j}(x')\label{eq:p5}
\end{multline}
 The   components of $\br g(\br x)$ in unprimed and primed systems
are related by
\begin{equation}
g'_{ij}(x')=\sum_{k=1}^{m}\sum_{l=1}^{m}g_{kl}(x)\dfrac{\partial x{}^{k}}{\partial x'^{i}}\dfrac{\partial x{}^{l}}{\partial x'{}^{j}}\label{eq:p6}
\end{equation}

Riemannian manifolds $(\ca M,\br g)$ use such a second rank covariant
tensor field to define a metric.\footnote{The term Riemannian manifold is assumed to include semi-Riemannian
manifolds.} The invariant inner product of two vector fields is a manifold object
defined as 
\begin{equation}
\Bigl\langle\br U(\br x),\br V(\br x)\Bigr\rangle=\br g(\br x)\bigl\{\br U(\br x),\br V(\br x)\bigr\}\label{eq:p7}
\end{equation}

Other forms and tensors are defined similarly. The important point
is that passive diffeomorphisms do not change manifold objects, but
only change the local coordinates and components used to represent
them. For this reason, differential geometric models in theoretical
physics universally assign reality only to manifold objects. Since
they do not change these manifold objects, passive diffeomorphisms
therefore do not change the physical reality being modeled. A magnetic
field is not identified with its components $B^{i}(x)$ or $B'^{i}(x')$.
It is modeled by a manifold object $\br B(\br x)$ that does not change
when the local coordinate system is changed. 

\section{Active Diffeomorphisms}

\label{sec:active}Active diffeomorphisms are the opposite of passive
diffeomorphisms. Passive diffeomorphisms change the system of   local
coordinates, but do not change the manifold objects. Active diffeomorphisms
change the manifold objects but do not change the system of   local
coordinates being used to represent those manifold objects.

Active diffeomorphisms are a special case of a more general differential
geometric construction, the differentiable mapping $\phi:\ca M\rightarrow\ca N$
from manifold $\ca M$ to manifold $\ca N$. In general, the two manifolds
may have different dimensions, and the mapping need not have an inverse. 

An active diffeomorphism is defined to be a differentiable, invertible
mapping from $\ca M$ to itself rather than to some other manifold
$\op N$, with the target manifold therefore having the same dimension
and the same system of   local coordinates.\footnote{This paper uses the term \emph{active diffeomorphism,} from \citet{Stachel-active}.
Einstein evidently called them what translates as \emph{point transformations.}
Differential geometric texts not aimed specifically at the general
relativity community do not even discuss active diffeomorphisms, limiting
themselves to the general case with $\phi:\ca M\rightarrow\ca N$
transforming between different manifolds. Several texts for general
relativists refer to active diffeomorphisms as simply \emph{diffeomorphisms
}with no modifier. We will always precede the word diffeomorphism
with a modifier, passive or active, to avoid confusion as to which
one is intended.} 

The mapping is $\phi:\ca M\rightarrow\ca M$ with manifold points
$\br x$ and $\tilde{\br x}$ before and after the mapping related
by $\tilde{\br x}=\phi(\br x)$. In terms of   local coordinates $\tilde{x}=\psi(\tilde{\br x})$
and $x=\psi(\br x)$, the active diffeomorphism is\footnote{Objects after transformation by an active diffeomorphism will be denoted
by a tilde over them, for example $\tilde{\br x}$ and $\tilde{x}$.}
\begin{equation}
\tilde{x}=\psi\circ\phi\circ\psi^{-1}(x)\label{eq:a1}
\end{equation}
The mapping is defined to be diffeomorphic; both \prettyref{eq:a1}
and its inverse are continuously differentiable to arbitrary order. 

Note that the transformed local coordinate values $\tilde{x}=\left(\tilde{x}^{1},\ldots,\tilde{x}^{m}\right)=\psi(\tilde{\br x})$
differ from the original local coordinate values $x=(x^{1},\ldots,x^{m})=\psi(\br x)$.
This is not because of a change of the local coordinate \emph{system}
but because the represented manifold object $\br x$ has been transformed
to a new manifold object $\tilde{\br x}=\phi(\br x)$. The unchanged
local coordinate \emph{system} is defined by the same homeomorphic
mapping $\psi$ before and after an active diffeomorphism. 

If manifold object $\br f(\br x)$ is a smooth function defined on
$\ca M$, then there is an actively transformed smooth manifold object
$\:\tilde{\br f}=\br f\circ\phi^{-1}$ with
\begin{equation}
\tilde{\br f}(\tilde{\br x})=\br f(\br x)\label{eq:a2}
\end{equation}
 called the \emph{push-forward }of $\br f$ by $\phi$ and denoted
$\tilde{\br f}=\phi_{*}\br f$. The original function $\br f$ is
also called the \emph{pull-back }of $\tilde{\br f}$ by $\phi$, and
is denoted $\br f=\phi^{*}\tilde{\br f}$. In terms of   local coordinates,
the relation between these functions is given by $\tilde{f}(\tilde{x})=f(x)$.

Tangent vectors can also be pushed forward or pulled back.\footnote{Unlike the general case where the mapping $\phi$ possibly had no
inverse and hence some relations were ill defined, active diffeomorphisms
have inverses and hence both pull-back and push-forward are always
well defined.} The push-forward $\tilde{\br V}=\phi_{*}\br V$, or equivalently
the pull-back $\br V=\phi^{*}\tilde{\br V}$, is defined by
\begin{equation}
\tilde{\br V}(\tilde{\br x})\,\tilde{\br f}(\tilde{\br x})=\br V(\br x)\,\br f(\br x)\label{eq:a3}
\end{equation}
In terms of   local coordinates, $\tilde{V}(\tilde{x})\tilde{f}(\tilde{x})=V(x)f(x)$.
The   component transformation, here written as a push-forward, is
\begin{equation}
\tilde{V}{}^{i}(\tilde{x})=\sum_{j=1}^{m}\dfrac{\partial\tilde{x}{}^{i}}{\partial x^{j}}V^{j}(x)\label{eq:a4}
\end{equation}

In Riemannian manifolds the metric tensor also can be equivalently
pushed forward $\tilde{\br g}=\phi_{*}\br g$ or pulled back $\br g=\phi^{*}\tilde{\br g}$.
The definition is 
\begin{equation}
\tilde{\br g}(\tilde{\br x})\bigl\{\tilde{\br V}(\tilde{\br x}),\tilde{\br W}(\tilde{\br x})\bigr\}=\br g(\br x)\bigl\{\br V(\br x),\br W(\br x)\bigr\}\label{eq:a5}
\end{equation}
for any general pair of tangent vectors. The component relation, here
expressed as a push-forward, is 
\begin{equation}
\tilde{g}_{ij}(\tilde{x})=\sum_{k=1}^{m}\sum_{l=1}^{m}g{}_{kl}(x)\dfrac{\partial x{}^{k}}{\partial\tilde{x}^{i}}\dfrac{\partial x{}^{l}}{\partial\tilde{x}^{j}}\label{eq:a6}
\end{equation}

In pre-general-relativistic differential geometry, a Riemannian metric
is a fixed part of the definition of a Riemannian manifold, denoted
$(\ca M,\br g)$. But the metric $\phi_{*}\br g$ pushed forward by
an active diffeomorphism could possibly be different from that fixed
metric of the manifold. Thus, in pre-general-relativistic physics,
in order to preserve general covariance it is necessary to restrict
active diffeomorphisms to those that do not change the Riemannian
metric,\footnote{See \citet{lee-riemann}, page 24. For example, in three-dimensional
models with a fixed Euclidean metric, the only permissible active
diffeomorphisms are rigid translations and rotations. A four-dimensional
model with a fixed Minkowski metric permits only rigid translations,
rotations, and boosts.} those with $\phi_{*}\br g=\br g$. Such active diffeomorphisms are
called \emph{isometric.}\footnote{In terms of   local coordinates and components, an isometric active
diffeomorphism has $\tilde{g}{}_{ij}(\tilde{x})=g_{ij}(\tilde{x})$.}\emph{ }In general relativity, however, the metric is unknown until
the Einstein field equation is solved. So there is no need to restrict
active diffeomorphisms in this way. In general relativity, the active
diffeomorphism is of the form $\phi:(\ca M,\br g)\rightarrow(\ca M,\phi_{*}\br g)$
with no isometric restriction on $\phi$.

\subsection{Generation of Active Diffeomorphisms}

\label{sec:generation}Some of the machinery of Lie Group theory can
be borrowed to generate active diffeomorphisms from given tangent
vector fields. A useful class of   active diffeomorphisms can be constructed
by considering the mapping $\phi_{\tau}$ along a given tangent vector
field $V(x)$.\footnote{Section 39 of \citet{arnold}, pages 68-70 and Chapter 9 of \citet{lee-smooth},
and pages 27-32 and 250-251 of \citet{oneill}.} 

Given a chosen starting point $\br x\in\ca M$, let a smooth mapping
$(0,\tau{}_{1})\rightarrow\ca M$ define a curve $\br x(\tau)$ starting
from $\br x(0)=\br x$. In terms of   local coordinates, this is $x(\tau)=\psi(\br x(\tau))$.
Differentiating this curve with respect to $\tau$ gives what is sometimes
called a \textquotedbl{}velocity\textquotedbl{} tangent vector, whose
components are $\dot{x}^{i}(\tau)=dx^{i}(\tau)/d\tau$. Given a general
tangent vector field $V(x),$ a curve whose velocity matches that
tangent vector for every $\tau\in(0,\tau_{1})$ is defined by the
set of differential equations 
\begin{equation}
\dot{x}^{i}(\tau)=V^{i}(x(\tau))\quad\quad\mbox{where}\quad\quad i=1,\ldots,m\label{eq:a7}
\end{equation}
whose solution $x(\tau)$ can be described as an integral curve or
\textquotedbl{}field line\textquotedbl{} of $V(x)$ passing through
$x(0)=x$. The corresponding field line in the manifold is then $\br x(\tau)=\psi^{-1}(x(\tau))$.

Since the tangent vector field is assumed to be defined at all points
of $\ca M$, we can consider the family of all such field-line curves
beginning at every point $\br x\in\ca M$. Consider an active diffeomorphic
mapping $\phi_{\tau}:\ca M\rightarrow\ca M$ which \emph{simultaneously}
carries each $\br x$ in $\ca M$ into an $\br x(\tau)$ along the
particular field line starting at that $\br x$. When $\tau=0$, this
mapping is the identity mapping $\phi_{0}=I$. When $\tau>0$, mapping
$\phi_{\tau}$ will move each point $\br x$ of $\ca M$ along the
appropriate field line to a new point $\tilde{\br x}=\br x(\tau)=\phi_{\tau}(\br x)$.
Expressing the same mapping in   local coordinates, each point $x=\psi(\br x)$
is moved by active diffeomorphism $\theta_{\tau}=\psi\circ\phi_{\tau}\circ\psi^{-1}$
into a new point $\tilde{x}=x(\tau)=\theta_{\tau}(x)$. 

If $V(x)$ is a so-called Killing Vector Field,\footnote{\citet{lee-smooth}, page 345.}
then by definition the active diffeomorphism $\phi_{\tau}$ is isometric.
Generation of more general active diffeomorphisms with $\tilde{\br g}=\phi_{*}\br g\neq\br g$
requires that $V(x)$ not be a Killing Vector Field.

\subsection{Associated Passive Diffeomorphisms}

\label{subsec:assoc}An active diffeomorphism transforms manifold
point $\br x$ with local coordinate values $x=\psi(\br x)$ to a
new manifold point $\tilde{\br x}=\phi(\br x)$ with new   local coordinate
values $\tilde{x}=\psi(\tilde{\br x})$. For each active diffeomorphism
there is an \emph{associated} passive diffeomorphism. If the active
diffeomorphism is applied first, and then the associated passive diffeomorphism
is applied, the end result is that the local coordinates are returned
to their original numerical values. The associated passive diffeomorphism
does not change the manifold point $\tilde{\br x}$, but undoes the
change of   local coordinate values produced by the active diffeomorphism.
It follows from \prettyref{eq:p1a} and \prettyref{eq:a1} that the
required associated passive diffeomorphism is $\tilde{x}''=\psi''\circ\psi^{-1}(\tilde{x})$
where $\psi''=\psi\circ\phi^{-1}$. When it is used as described above,
the final result is $\tilde{x}''=x$.

It may seem that after this sequence nothing has changed. The   local
coordinates are back to their original values. Examination of the
transformation rules for functions, tangent vectors, and general tensors
shows that after the above sequence, the final components of these
coordinate objects are also returned to their original numerical values.
However, the correct reading is not that \emph{nothing} has changed
but that \emph{everything} has changed, both the manifold objects
and the system of local coordinates used to represent them. Due to
general covariance, everything has changed in a consistent manner. 

\section{Uncontested Points}

\label{sec:summary}Newton equivalence and Leibniz equivalence are
two different interpretations of active diffeomorphisms and their
use in physics. But the two interpretations agree on the following
points.
\begin{enumerate}
\item Coordinate objects change when the local coordinate system is changed
by a passive diffeomorphism (a diffeomorphic change of   local coordinates);
manifold objects do not. Therefore, differential geometric models
in theoretical physics universally assign reality only to manifold
objects.
\item Passive diffeomorphisms do not change manifold objects. Therefore
passive diffeomorphisms do not change the physical reality being modeled.
\item Active diffeomorphisms change manifold objects. Whether this change
is a true change of the physical situation being modeled is a point
of contention between the two interpretations. But both agree that
the manifold objects are modified by active diffeomorphisms.
\item Let a generally covariant model of an experiment consist of the set
of manifold objects $\br A=(\br a_{1},\ldots,\br a_{n})$ and the
outcome of the experiment consist of the set of manifold objects $\br B=(\br b_{1},\ldots,\br b_{k})$.
If an active diffeomorphism is performed on the model, the result
will be a model consisting of transformed manifold objects $\tilde{\br A}=(\tilde{\br a}_{1},\ldots,\tilde{\br a}_{n})$
with the transformed outcome manifold objects $\tilde{\br B}=(\tilde{\br b}_{1},\ldots,\tilde{\br b}_{k})$.
If $\br A$ produces $\br B$, then $\tilde{\br A}$ produces $\tilde{\br B}$.
Colloquially speaking, due to general covariance nothing is \textquotedbl{}broken\textquotedbl{}
by the active diffeomorphism and the model will still \textquotedbl{}work\textquotedbl{}
as before the active diffeomorphism was done.
\item As described in \prettyref{subsec:assoc}, any active diffeomorphism
can be followed by an associated passive diffeomorphism that returns
the local coordinates and components to their original numerical values.
\end{enumerate}

\section{Two Examples of Active Diffeomorphisms}

\label{sec:examples}Discussion of abstract issues like the role of
active diffeomorphisms in differential geometry is often aided by
concrete examples of what the abstractions actually produce. Sections
\ref{sec:newt} and \ref{sec:leibniz} will make reference to the
following two examples showing the effect of active diffeomorphisms
on simple physical models.

\subsection{Example 1}

\begin{figure}[h]
\centerline{\includegraphics[scale=0.2]{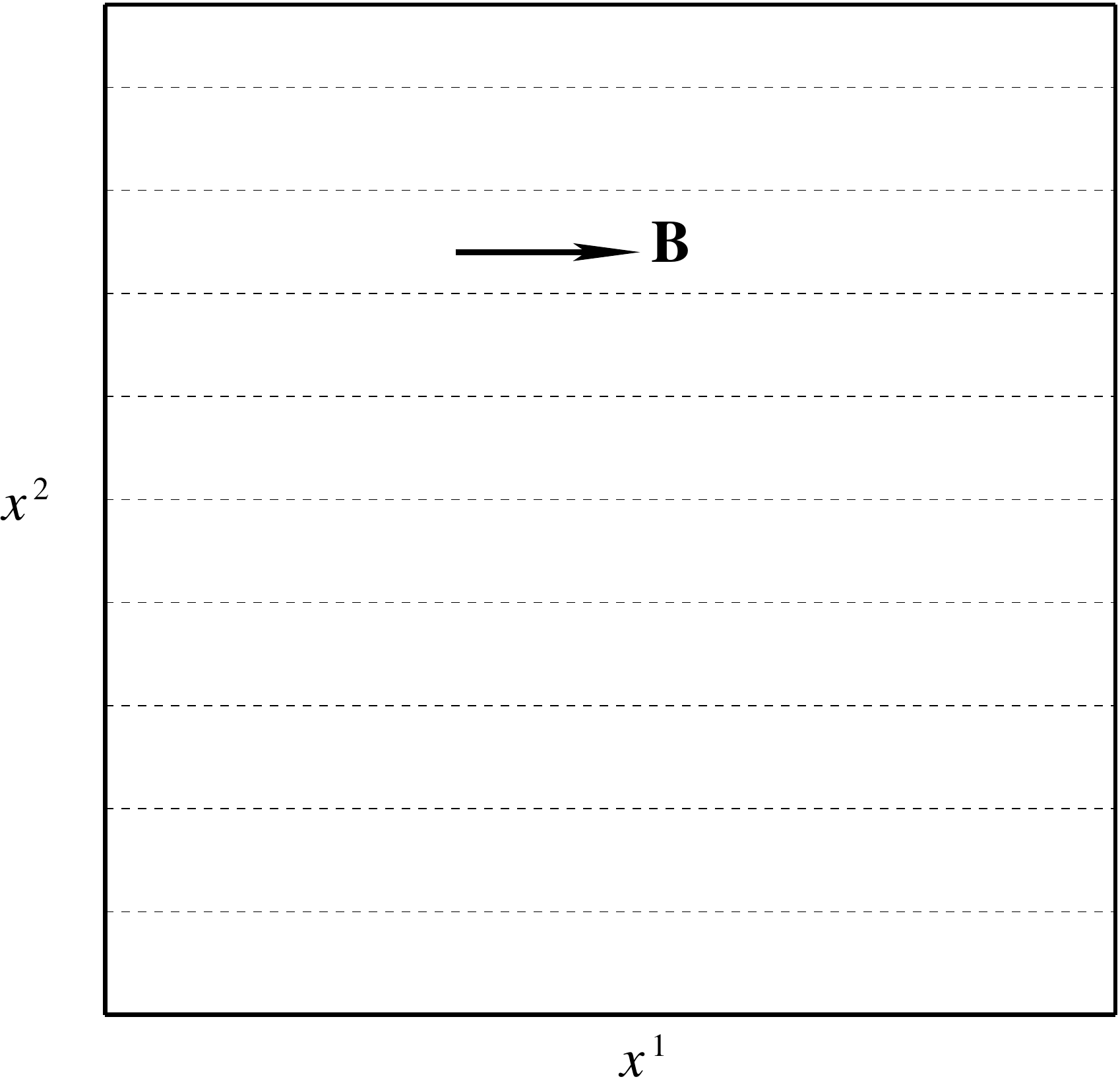}$\quad\quad$\includegraphics[scale=0.2]{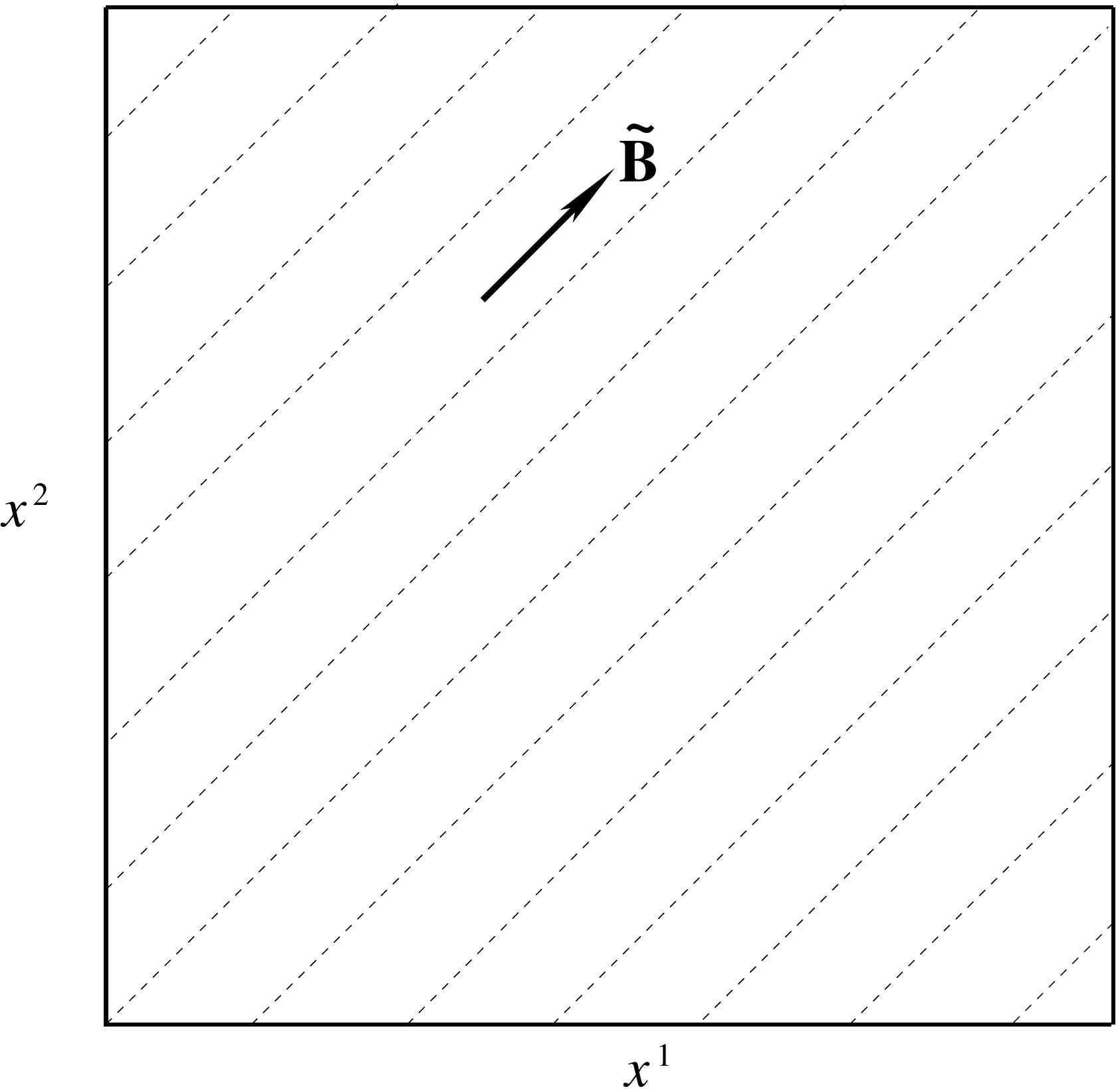}$\quad$$\quad$\includegraphics[scale=0.2]{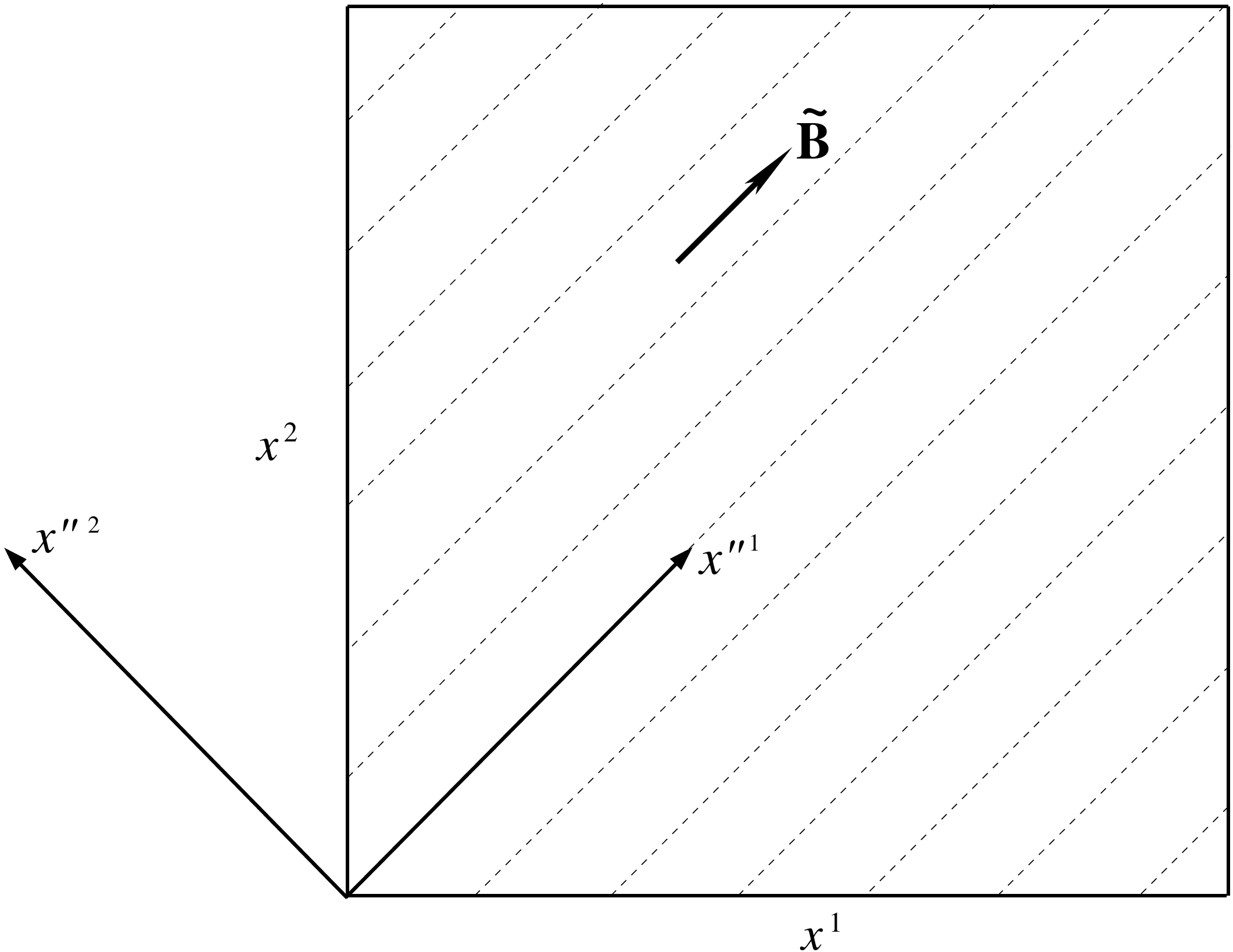}}

\caption{a, b, and c.\label{fig:5.1}}
\end{figure}

\label{subsec:ex1}\prettyref{fig:5.1}a shows a model of an experiment
containing a uniform magnetic field $\br B(\br x)$ in a three-dimensional
Euclidean space,\footnote{Or, equivalently, Examples 1 and 2, and the example in \prettyref{subsec:Leibniz-extrapolation},
show time slices of limited, approximately Minkowskian, regions of
spacetime. See Section 10.3 of \citet{rindler}.} viewed from a system of   local coordinates in which the magnetic
field has the components $(B^{1},0,0)$. The field lines are shown
as dashed lines, and a typical field value is shown as an arrow. An
active diffeomorphism is generated using the methods of \prettyref{sec:generation}
using a tangent vector field $V(x)$ with components $(0,-x^{2},x^{1},0)$.
This vector field generates an orthogonal rotation by angle $\tau$.
This active diffeomorphism is assumed to be the identity everywhere
except in the apparatus. Applying this active diffeomorphism with
$\tau=\pi/4$ produces \prettyref{fig:5.1}b. The local coordinate
system (possibly defined by the edges of the table) is unchanged,
but the field lines are now at a $45^{\circ}$ angle. A typical actively
transformed field value $\tilde{\br B}$ is shown. \prettyref{fig:5.1}c
is the same as \prettyref{fig:5.1}b, except that the associated passive
diffeomorphism has now been added, giving the local coordinate axis
lines shown as $x''^{1}$ and $x''^{2}.$ The actively transformed
field lines are in the direction of the $x''^{1}$ axis, as described
in \prettyref{subsec:assoc}, and $\tilde{B}''^{i}=B^{i}$ for $i=1,2,3$.

\subsection{Example 2}

\begin{figure}[h]
\centerline{\includegraphics[bb=0bp 0bp 460bp 459bp,clip,scale=0.265]{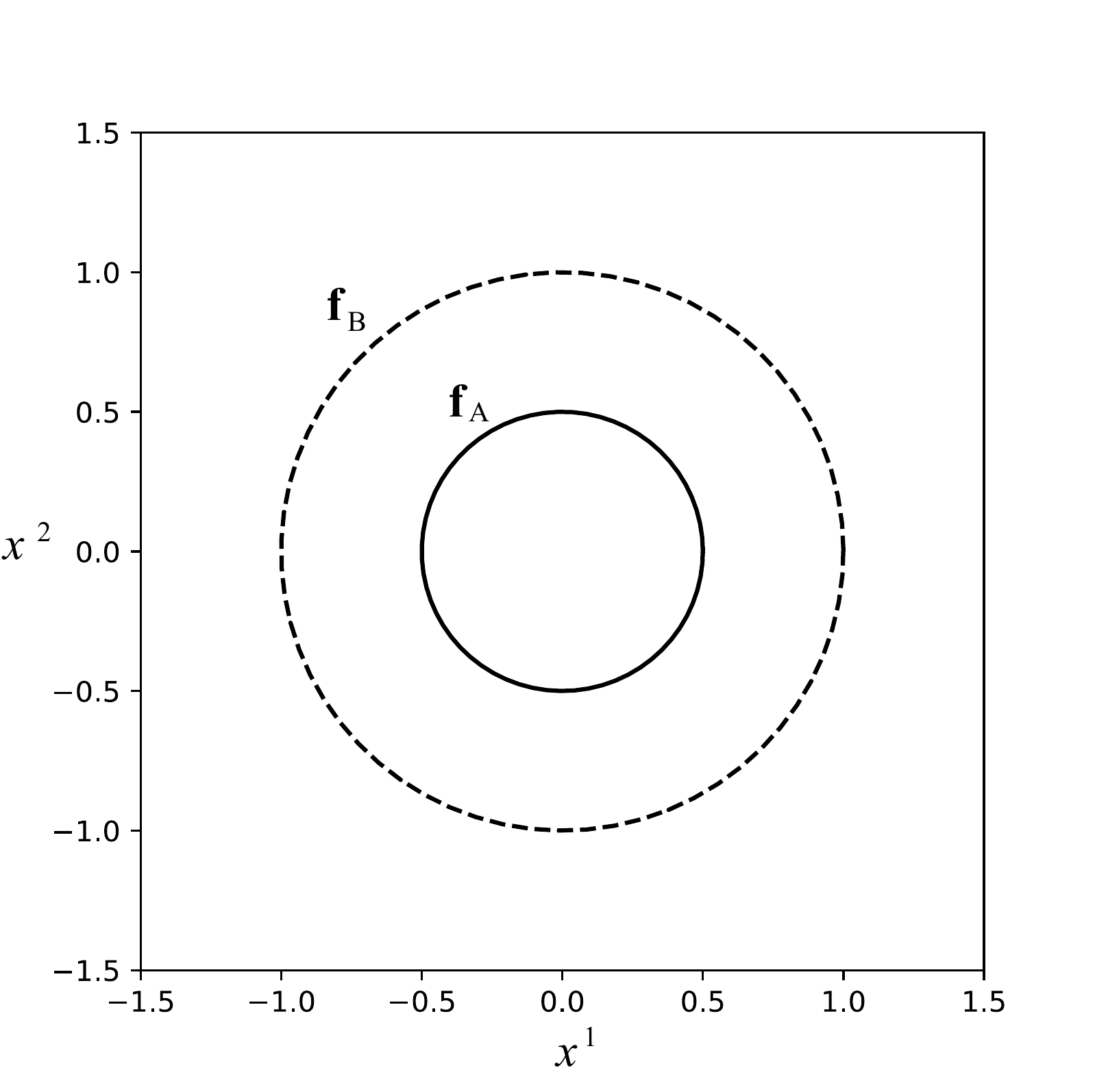}\includegraphics[clip,scale=0.26]{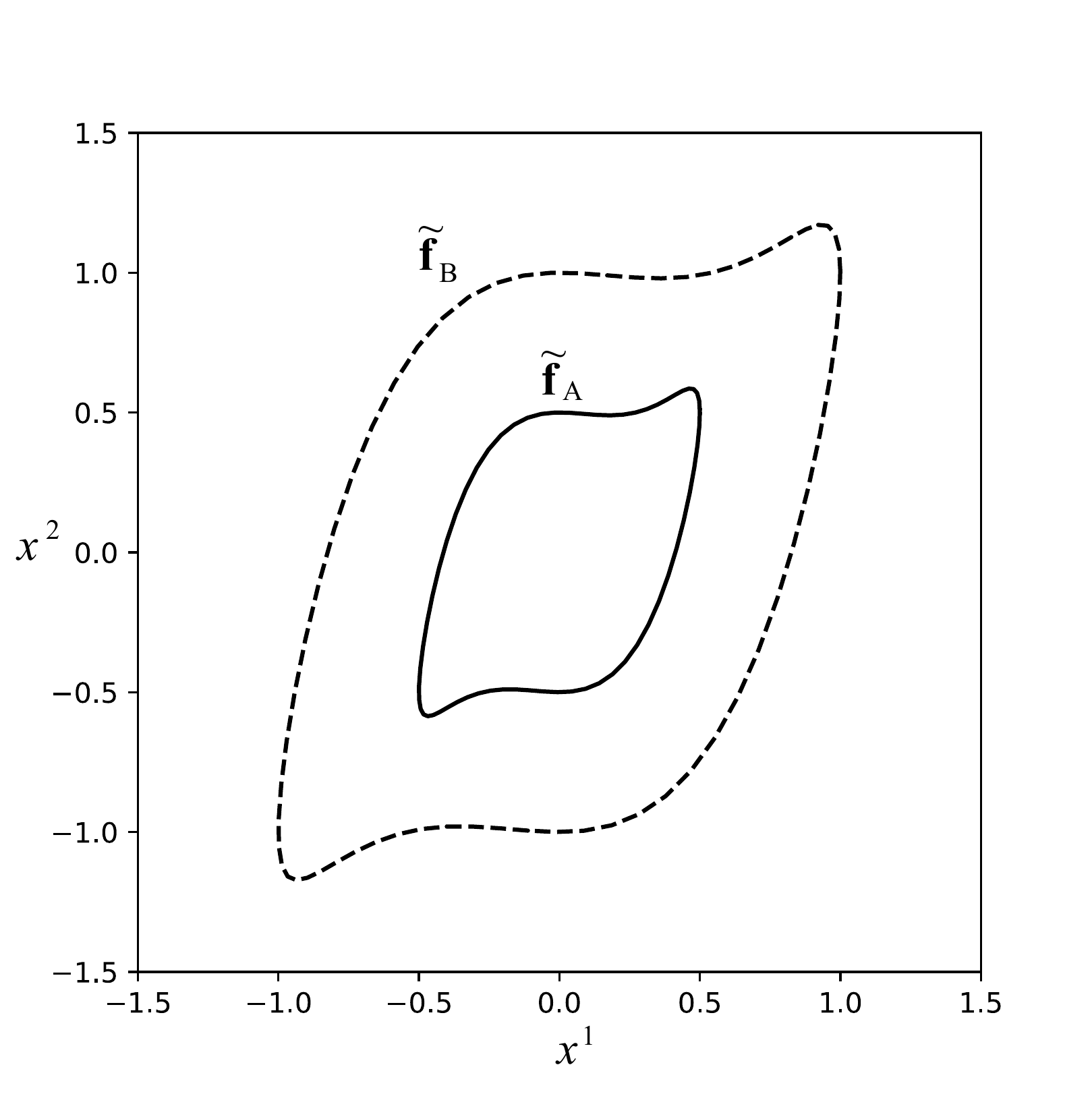}\includegraphics[bb=0bp 0bp 466bp 467bp,clip,scale=0.24]{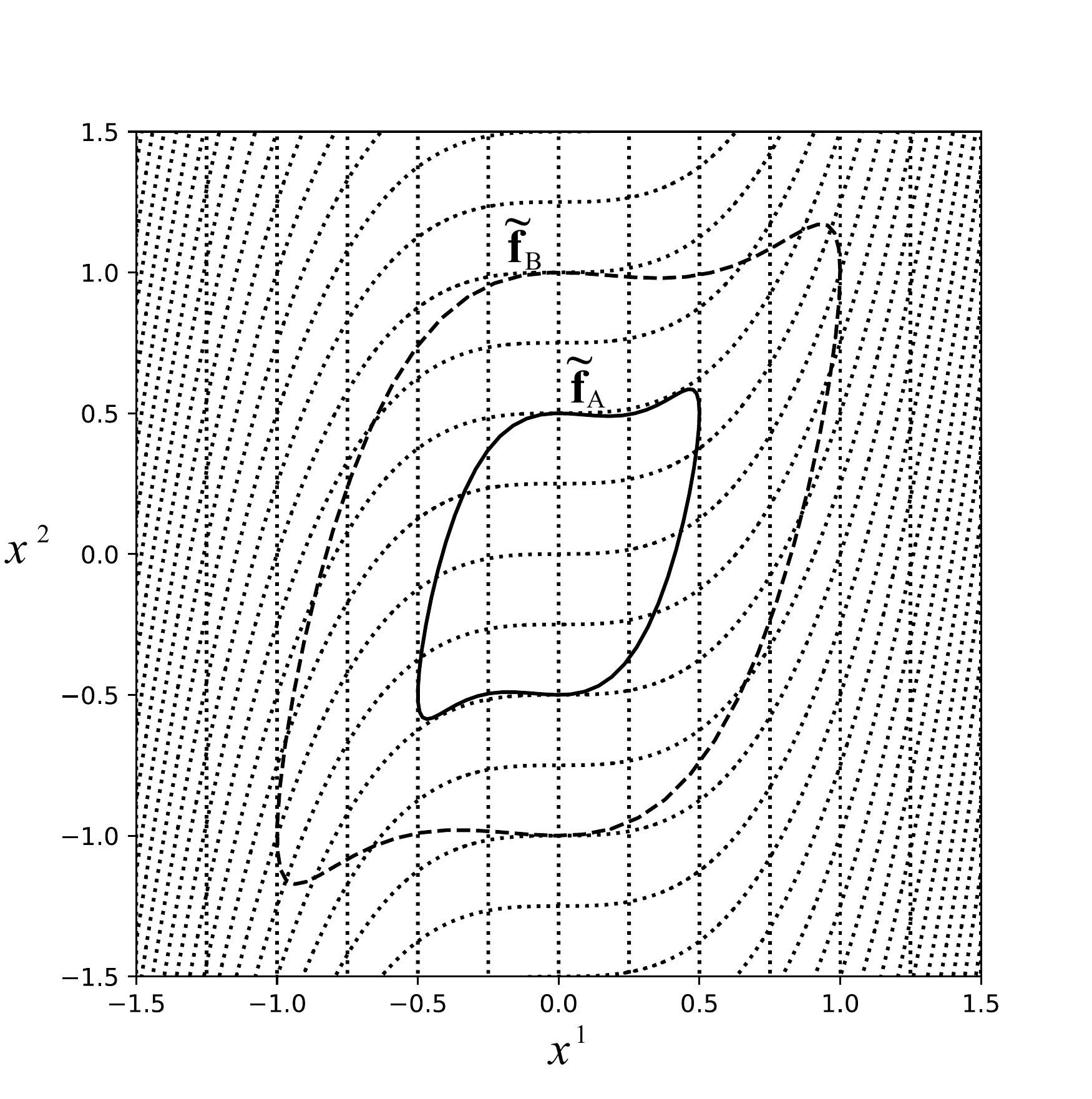}}

\caption{a, b, and c.\label{fig:5.2}}
\end{figure}

\label{subsec:ex2}\prettyref{fig:5.2}a shows a model in which a
mass distribution, as viewed initially on a hyperplane at time $t_{A}$,
has a circular cylindrical boundary $\br f_{\text{A}}$. It later,
as viewed on a hyperplane at time $t_{B}$, evolves into a circular
cylindrical distribution with larger boundary $\br f_{\text{B}}$.
An active diffeomorphism is generated using the methods of \prettyref{sec:generation}
using the tangent vector field $V(x)$ with components $\bigl(0,0,\bigl(x^{1}\bigr)^{3}\!\!,0\bigr)$.
Applying this active diffeomorphism with $\tau=1$ to the model gives
\prettyref{fig:5.2}b in which the mass boundaries are distorted.
But, as noted in point 4 of \prettyref{sec:summary}, due to general
covariance the actively transformed initial boundary $\tilde{\br f}_{\text{A}}$
will still evolve into the actively transformed final boundary $\tilde{\br f}_{\text{B}}$.
Note that the distortion in \prettyref{fig:5.2}b is due to distortion
of the manifold object itself; the local coordinate system is unchanged.
\prettyref{fig:5.2}c shows the grid lines of the associated reference
system $x''$. In this distorted reference system, the actively transformed
mass distributions would have the same local coordinate description
as the original distributions had in \prettyref{fig:5.2}a. 

\section{Newton Equivalence}

\label{sec:newt}Newton Equivalence holds that active diffeomorphisms
that change manifold objects necessarily change the physical reality
being modeled by those objects. However, due to general covariance,
the different physical situations reached by active diffeomorphisms
are all equally possible. They model different but equally possible
physical situations.

Thus, in Example 1 above, \prettyref{fig:5.1}b with its rotated $\tilde{\br B}$
field models a physically modified experiment. This accords with the
usual practice in experimental physics. If an experimenter finds that,
with no change in the system of local coordinates, the magnetic field
is now pointing in a different direction, his immediate conclusion
is that something physical has changed. That the experiment still
behaves correctly is taken as evidence for invariance under rotations.
The fact that a rotation of reference system to match the changed
magnetic field, as in \prettyref{fig:5.1}c, would return the local
coordinates of that field to their original values would be taken
as further evidence of rotational symmetry, and would not change the
conclusion about \prettyref{fig:5.1}b.

In Example 2 above, an astrophysicist observes the mass distribution
in \prettyref{fig:5.2}a to evolve into a larger one of the same circular
shape. Her generally covariant model of this phenomenon can be transformed
by an active diffeomorphism into a model that predicts the evolution
of a distorted mass distribution into a distorted larger one, as in
\prettyref{fig:5.2}b. However, in the model or in actual observation
of the phenomena, the astrophysicist would never consider the two
figures to represent the same physical reality. \prettyref{fig:5.2}b
represents a different physical situation, but one that could happen
given the validity of the generally covariant model. 

\subsection{Symmetry and Identity}

\label{subsec:newton-symmetry}An argument for Newton Equivalence
is its agreement with the treatment of symmetry in physics. Active
diffeomorphisms are a kind of symmetry operation allowed by generally
covariant physical models.\footnote{A generally covariant model is one constructed using only manifold
objects. Also, equalities are only between manifold objects of the
same tensorial character: scalar equals scalar, vector equals vector,
second rank covariant tensor equals second rank covariant tensor,
etc. Thus an active diffeomorphism transforms a generally covariant
model into another generally covariant model. } The actively transformed version of a successful model is another
successful model, something that could also happen. But the transformed
model is not identical to the original one; it models a different
experiment that is also possible, related to the original one by the
symmetry but not identical to it.

In theoretical and experimental physics, great care is taken to distinguish
\emph{symmetry} from \emph{identity. }This distinction is particularly
emphasized in discussion of discrete symmetries like parity and time
reversal. Parity symmetry means that the parity transformed experiment
is one that is also possible, \emph{i.e.,} is not prohibited by the
laws of physics. But it does not mean that original and transformed
experiments are identical, are physically the same. They are different.
Parity symmetry only means that the parity transformed experiment
is one that can happen without violating the laws of nature.\footnote{For example, Section 15.11 of \citet{bjorken-drell} says of the active
parity transformation, \textquotedbl{}If we invert the measuring apparatus,
that is, consider a new physical system ... the dynamics of the new
system is the same as that of the original one, provided parity is
conserved.\textquotedbl{}} 

The distinction between symmetry and identity is even more pointed
in time reversal invariance. Consider an experiment in which a particle
fissions into two daughter particles. Time reversal symmetry means
that the inverse process, two particles coalescing into one, is something
that could possibly happen. The laws of physics do not prohibit it.
But it does not mean that this inverse process is physically the same
as the original one. It is different, and the experiment to demonstrate
the time reversed process could be very difficult to perform.

Newton Equivalence echos this strong distinction between symmetry
and identity found in theoretical physics. In a generally covariant
model, an active diffeomorphism is a symmetry operation that changes
the physical situation being modeled. The transformed model is then
equally possible, something not prohibited by the generally diffeomorphic
laws of nature.

\subsection{Observer and Apparatus}

\label{subsec:newton-observer}Another argument for Newton Equivalence
is its agreement with the general usage of coordinate objects and
manifold objects in theoretical physics. The application of differential
geometry to physics is based on a distinction between observer and
apparatus. Local coordinates are the numbers resulting from the observation
of an apparatus. The possibility of observation of an experiment from
a viewpoint outside that experiment is fundamental to this distinction.
Since they modify manifold objects, active diffeomorphisms do change
the \emph{values} of the data collected from an experiment, but they
do not transform the \emph{system} used by the observer to extract
that data. 

For example, let the magnetic field in \prettyref{fig:5.1} be from
an electromagnet sitting on a laboratory table, with a system of local
coordinates fixed to the table. If that electromagnet apparatus is
rotated by 45 degrees relative to the laboratory table, the result
is a physical change, as in \prettyref{fig:5.1}b. The \emph{system}
of local coordinates is still fixed to the table, but the \emph{values}
of the local coordinates have changed, due to the physical rotation
of the apparatus.

In the Newton Equivalence interpretation, the apparatus is modeled
by coordinate independent manifold objects and the observation is
modeled by local coordinates. Seeing the local components of the magnetic
field change from $B$ to $\tilde{B}$ with no change in the system
of local coordinates, the observer correctly concludes that a physical
change has taken place. This use of active and passive diffeomorphisms
in differential geometry echos the usual practice in  physics.

\subsection{Agnosticism}

The name \emph{Newton} Equivalence was chosen to suggest a tenable
interpretation of active diffeomorphisms that is also consistent with
current practice in physics. It is not intended to imply that an adherent
of this interpretation must necessarily take Newton's side in the
Newton-Leibniz debates of the early eighteenth century.\footnote{See \emph{The Leibniz-Clarke Correspondence,} \citet{LeibnizClarke}.}
As discussed in \prettyref{sec:substantivalism} below, use of the
Newton Equivalence interpretation allows a researcher to remain agnostic
about the substantivalist-relativist debate.

\section{Leibniz Equivalence}

\label{sec:leibniz}Leibniz Equivalence states that, \textquotedbl{}Diffeomorphic
models represent the same physical situation.\textquotedbl{}\footnote{Earman and Norton, \emph{op. cit., }page 522. As noted in \prettyref{sec:intro},
their term \textquotedbl{}diffeomorphic models\textquotedbl{} is to
be interpreted as models related by active diffeomorphisms.} While an active diffeomorphism does change the manifold objects of
a model, it is asserted not to change the \textquotedbl{}physical
situation\textquotedbl{} being modeled by those manifold objects.
Thus the physical situation is not identified with a particular set
of manifold object values, but with an \emph{equivalence class} of
manifold object values, a given set of values together with all the
others reached from it by application of any active diffeomorphism.
\prettyref{fig:5.1}a and \prettyref{fig:5.1}b thus are asserted
to show the same physical situation. They are not models of a different
physical reality; rather they are just parts of the same physical
reality. 

Some support for Leibniz Equivalence comes from its resonance with
Leibniz' objections to Newton's absolute space. If we imagine the
experiment in \prettyref{fig:5.1} to be mounted in a closed room
sitting on a turntable, then if someone rotates the turntable in the
night while the experimenter sleeps, we would go directly from \prettyref{fig:5.1}a
to \prettyref{fig:5.1}c and the experimenter would not discern any
changes to his experiment. In this case, the interior of the room
is like the experimenter's Leibnizian universe and the rest of the
Earth is like Newton's absolute space. But the analogy is flawed;
there are discernible changes in the physical situation. For example,
the workers who rotated the laboratory in the night would have observed
the change, and would have carefully measured it to be $45^{\circ}$.
Also, \prettyref{fig:5.1}b would still require additional justification,
since in it only the apparatus is rotated, and the unchanged laboratory
reference system detects this change.

Leibniz Equivalence also holds \prettyref{fig:5.2}a and \prettyref{fig:5.2}b
to show the same physical situation. Here the difference is not one
of a Leibnizian rigid translation of a whole universe. It is a distortion,
of a sort that neither Newton nor Leibniz would be likely to have
imagined. The Leibniz interpretation ignores the large variety of
available active diffeomorphisms; active diffeomorphisms can be much
more general than the simple displacements argued by Leibniz.

\subsection{Symmetry and Identity}

\label{subsec:Leibniz-symmetry}Leibniz Equivalence erases the difference
between symmetry and identity, the concepts that theoretical physics
seeks to keep distinct. The name of the Leibniz Equivalence principle
uses the word \textquotedbl{}equivalence\textquotedbl{} but the principle
itself asserts identity, the \textquotedbl{}same\textquotedbl{} physical
situation. As noted in \prettyref{subsec:newton-symmetry} above,
general covariance in differential geometry is a kind of symmetry
principle; given general covariance, applying an active diffeomorphism
to a successful model produces another model that is equally successful.
In theoretical physics, such symmetry transformations are carefully
distinguished from identity. Leibniz Equivalence removes this distinction;
active diffeomorphisms merely move the model from one to another member
of an equivalence class, with both members representing the identical
reality.

Thus, referring to \prettyref{subsec:newton-symmetry} above, Leibniz
Equivalence would say that an electromagnet sitting on a laboratory
table, and the same electromagnet after the active diffeomorphism
symmetry operation of rotation by 45 degrees relative to the table,
both represent the identical \emph{physical situation. }This interpretation
diverges from the standard meanings of symmetry and \emph{physical
situation} in physics.

\subsection{Passive and Active}

\label{subsec:Leibniz-passive-active}Leibniz Equivalence blurs the
difference between passive and active diffeomorphisms. As noted in
items 1 and 2 of \prettyref{sec:examples}, it is generally accepted
that \emph{passive} diffeomorphisms do not change the physical reality
being modeled. This is because they do not change manifold objects,
which are used exclusively to model real quantities. Leibniz Equivalence
extends this behaviour to \emph{active} diffeomorphisms, even though
active diffeomorphisms do change manifold objects. It says that different
manifold objects produced by active diffeomorphisms \textquotedbl{}represent\textquotedbl{}
the same physical situation. But, unlike coordinate objects which
do merely \textquotedbl{}represent\textquotedbl{} an underlying manifold
object in some arbitrary local coordinate system, the manifold objects
themselves do not \textquotedbl{}represent\textquotedbl{} some deeper
reality; the manifold objects \emph{are} the differential geometric
model of that reality. Changing them changes the reality being modeled.

\subsection{Observer and Apparatus}

\label{subsec:Leibniz-observer}In Section 4 of their paper, Earman
and Norton deny the theoretical physicist's apparatus-observer distinction
discussed in our \prettyref{subsec:newton-observer}. They assert:
\emph{To complete the dilemma we need only note that spatio-temporal
positions by themselves are not observable. Observables are a subset
of the relations between the structures defined on the spacetime manifold.
Thus we cannot observe that body b is centered at position x. What
we do observe are such things as the coincidence of body b with the
x mark on a ruler, which is itself another physical system. Thus observables
are unchanged under }{[}active{]}\emph{ diffeomorphism. Therefore
}{[}active{]}\emph{ diffeomorphic models are observationally indistinguishable.}\footnote{Earman and Norton, \emph{op. cit.,} page 522.}

If applied only to a particular case—Leibniz' rigid translation of
the entire universe relative to Newton's absolute space—this assertion
is a paraphrase of Leibniz' argument that the unobservability of a
rigid translation of the entire universe renders spatiotemporal position
relative to Newton's absolute space meaningless. However, the application
of the same argument to active diffeomorphisms is untrue in general.

Their concluding statement that \textquotedbl{}...observables are
unchanged under {[}active{]} diffeomorphism. Therefore {[}active{]}
diffeomorphic models are observationally indistinguishable\textquotedbl{}\footnote{By \textquotedbl{}diffeomorphic models,\textquotedbl{} Earman and
Norton mean models related by an active diffeomorphism.} is untrue for the general case of \emph{localized} active diffeomorphisms.\footnote{The term \emph{localized active diffeomorphism} will be used here
to denote active diffeomorphisms whose domain is the whole manifold,
but which act as the identity transformation outside of a specified
region of spacetime. Einstein's active diffeomorphisms that are the
identity except in a \textquotedbl{}hole\textquotedbl{} are an example.} \emph{Localized }active diffeomorphisms\emph{ }can transform one
part of the universe while leaving the rest of it unmodified, and
therefore are observable from a viewpoint in the unmodified region. 

To illustrate by a simple example, consider an apparatus sitting on
a laboratory table. A localized active diffeomorphism now spatially
translates the entire apparatus across the table to a new position,
but does not transform the table or the rest of the universe. Newton
Equivalence holds that this move across the table is a real, observable
physical change but, due to spatial translation symmetry, does not
upset the working of the apparatus. If any part of the apparatus initially
had local coordinate $x_{o}$ in an unchanged and fixed system of
  local coordinates (for example one defined by the table edges),
after the move it will now have a different local coordinate $\tilde{x}_{o}$.

The Earman-Norton statement says that, since this move of the apparatus
across the table is done by an active diffeomorphism, the two positions
of the apparatus must be \emph{observationally indistinguishable}.
But, relative to the untransformed table, both the initial position
of the apparatus, and its changed position after the localized active
diffeomorphism, are plainly observable and distinguishable. 

This example illustrates a general result. In the general case of
localized active diffeomorphisms there will always be a fixed reference
systems, anchored in the untransformed region, that will register
changes, and thus render an active diffeomorphism observable. It is
untrue in general that \textquotedbl{}...{[}active{]} diffeomorphic
models are observationally indistinguishable\textquotedbl{} as asserted
in the above quotation.

Another failure is that the above quotation's application of Leibnizian
relativity to active diffeomorphisms ignores complex distortions such
as that in \prettyref{fig:5.2} that cannot be reduced to Leibnizian
arguments about translation with an unchanged relation of internal
structures. 

In summary, the Earman-Norton assertion quoted at the beginning of
this subsection is an unjustified extrapolation, from Leibnizian relativism
in the particular case of the Newton-Leibniz debates, to all active
diffeomorphisms. It ignores the variety of available active diffeomorphisms,
and is untrue for the general case of localized active diffeomorphisms.

\subsection{History}

\label{subsec:Leibniz-history}The principal argument for Leibniz
Equivalence appears to be historical, that it generalizes Einstein's
resolution of his hole argument dilemma. Einstein maintained that
the different metric solutions to his field equation represent the
same physical reality. As expressed by Hawking and Ellis, \textquotedbl{}...
the model for spacetime is not just one pair $(\op M,\br g)$ but
a whole equivalence class of all pairs $(\op M',\br g')$ which are
equivalent to $(\op M,\br g)$.\textquotedbl{}\footnote{\citet{HawkingEllis}, page 56. }
Some general relativity texts also echo this assertion, but, notably,
others do not,\footnote{For example, \citet{misner-thorn-wheeler}, \citet{weinberg}, and
\citet{rindler}.} preferring to move directly from the field equation to its practical
applications to cosmology. 

It has been suggested by \citet{Johns-validity} that there may be
a less drastic escape from Einstein's dilemma. The local coordinates
used to write Einstein's field equation do not have any physical meaning
until \emph{after} a metric solution is found that defines their relation
to physical quantities. Therefore, it may be possible to reject as
spurious a metric solution that gives a physical meaning to its local
coordinates that violates a desired symmetry, for example spherical
symmetry for the Schwarzschild solution. In this way, as for any differential
equation, rejection of spurious solutions can lead to a unique solution.

Aside from the question of the necessity of Einstein's solution to
his hole dilemma is the question of Earman and Norton's generalization
of that solution to a principle covering virtually all uses of differential
geometry in physics. Einstein treated a specific example of a specific
theory. Generalization of Einstein's solution to the principle of
Leibniz Equivalence is by its nature an unprovable assertion. 

\subsection{Extrapolation}

\label{subsec:Leibniz-extrapolation}Earman and Norton assert without
further proof that \textquotedbl{}...{[}active{]} diffeomorphism is
the counterpart of Leibniz' replacement of all bodies in space in
such a way that their relative relations are preserved.\textquotedbl{}\footnote{Earman and Norton, \emph{op. cit.,} page 521.}
But active diffeomorphisms are far more general than the rigid translation
of the entire universe used by Leibniz. As noted in \prettyref{subsec:ex2},
they can distort the transformed objects rather than simply moving
them, thus upsetting their \textquotedbl{}relative relations.\textquotedbl{}
And, as discussed in \prettyref{subsec:Leibniz-observer}, they can
transform parts of the universe while remaining the identity in other
parts, thus permitting reference systems  that render the transformation
observable. The extrapolation of Leibnizian relativism to all active
diffeomorphisms is unjustified.

These extrapolations will appeal to a researcher approaching the subject
of differential geometry from a Leibnizian perspective. A researcher
viewing the world through a Leibnizian lens may simply take the principle
of Leibniz Equivalence as a \emph{definition} of the technical phrase
\emph{physical situation, }a phrase now referring to an equivalence
class of observationally distinct manifold objects related to each
other by active diffeomorphisms.  Then the Leibniz Equivalence principle
becomes a tautology. Also such a researcher may want to replace the
physics term \emph{unobservable} by the technical term \emph{operationally
indistinguishable} defined by the assertion quoted at the beginning
of\emph{ }\prettyref{subsec:Leibniz-observer}. As noted in \prettyref{subsec:Leibniz-observer},
this redefinition would define the \emph{observable} differences produced
by localized active diffeomorphisms to be \emph{observationally indistinguishable.}
With these redefinitions, the principle of Leibniz Equivalence, together
with the assertion quoted at the beginning of \prettyref{subsec:Leibniz-observer},
become true by definition.

But these extrapolations would produce an interpretation of differential
geometry that misunderstands the generality of active diffeomorphisms,
and is significantly at variance with the usages in current theoretical
and experimental physics. 

\section{Substantivalism}

\label{sec:substantivalism}In Section 3 of their paper,\footnote{Earman and Norton, \emph{op. cit.,} page 521.}
Earman and Norton state the \textquotedbl{}acid test\textquotedbl{}
of substantivalism to be the denial of the Leibnizian relativist proposition
that the whole universe displaced \textquotedbl{}three feet East\textquotedbl{}
is the same universe. The Earman-Norton arguments against substantivalism
rest on their assertion that a substantivalist, in denying this Leibnizian
relativism, must also deny their Leibniz Equivalence principle. Since
Leibniz Equivalence is an extrapolation of Leibnizian relativism,
this assertion is true. But this does \emph{not} imply the converse,
that one who denies Leibniz Equivalence must therefore deny Leibnizian
relativism and take Newton's side in the Newton-Leibniz debates.\footnote{Leibniz Equivalence (LE) implies Leibnizian relativism (LR). Therefore
substantivalism (not LR) implies (not LE). But the converse, that
(LR) implies (LE), is not provable. It is an extrapolation that Newton
Equivalence denies. Therefore (not LE) does not imply substantivalism
(not LR).} Instead, Newton Equivalence denies only Earman and Norton's \emph{extrapolation,}
from  Leibnizian relativism for the universe, to their principle of
Leibniz Equivalence for all active diffeomorphisms regardless of scale
and type. An adherent to Newton Equivalence may be, but need not be,
a substantivalist.

We now consider the two arguments that Earman-Norton make against
substantivalism. Each of them suggests an undesirable consequence
resulting from the denial of Leibniz Equivalence.

In their Section 4, titled \textquotedbl{}The Verificationist Dilemma,\textquotedbl{}\footnote{Earman and Norton, \emph{op. cit.,} page 522.}\emph{
}Earman and Norton argue that, since a substantivalist must deny Leibniz
Equivalence, he or she is committed to accept that the \emph{distinct
states} produced by active diffeomorphisms are physically different,
which runs counter to their argument, quoted at the beginning of our
\prettyref{subsec:Leibniz-observer}, that such differences are \emph{observationally
indistinguishable.} Newtonian Equivalence does indeed consider the
\emph{distinct states,} \emph{i.e.,} the different values of manifold
objects produced by active diffeomorphisms, to be observably different.
But, as discussed in \prettyref{subsec:Leibniz-observer}, the Earman-Norton
claim that the differences produced by any active diffeomorphism must
be unobservable is based on a misunderstanding of the generality of
available active diffeomorphisms, and is untrue in general. Thus Newton
Equivalence escapes the dilemma.

In their Section 5, titled \textquotedbl{}The Indeterminism Dilemma,\textquotedbl{}\footnote{Earman and Norton, \emph{op. cit.,} page 522.}
Earman and Norton argue that the substantivalist denial of Leibniz
Equivalence commits one to a \textquotedbl{}...radical local indeterminism.\textquotedbl{}\footnote{Earman and Norton, \emph{op. cit.,} page 524.}
But their argument fails because it misunderstands Einstein's proof
of indeterminism.

Einstein's proof requires a special experiment in which the energy-momentum
tensor source term $T_{\mu\nu}(x)$ in his field equation vanishes
in a \emph{hole region.} Then an active diffeomorphism is carefully
tailored to be the identity except in \emph{that same hole region
}where $T_{\mu\nu}(x)$ vanishes. This matched active diffeomorphism
then changes the solution without changing the source,\footnote{Outside the hole, the active diffeomorphism is the identity and hence
does not change the source there. Inside the hole, the source is identically
zero and hence is not transformed, since zero tensors transform to
zero tensors regardless of the active diffeomorphism applied.} and can therefore be used to prove indeterminacy of solution for
that particular thought experiment. The \emph{exact match} of active
diffeomorphism to source hole is essential to the argument. Without
it, the active diffeomorphism also modifies the source term and Einstein's
proof fails.\footnote{See Section 4 of \citet{Johns-validity}.} The
problem exposed by Einstein's hole argument is not multiple solutions
\emph{per se,} but multiple solutions \emph{all of which have the
same source; }specification of a source does not specify a unique
solution produced by that source.\footnote{See, for example, Einstein's 1913 letter to Ludwig Hopf, quoted on
page 163 of \citet{torretti}, in which he says, \textquotedbl{}It
is easily proved that a theory with generally covariant equations
cannot exist if we demand that the field be mathematically \emph{completely
determined by matter.}\textquotedbl{} (Italics mine.)} 

But Earman-Norton do not assume a match between a source and what
they call an active \textquotedbl{}hole diffeomorphism.\textquotedbl{}
Thus their unmatched active diffeomorphisms change both the solution
\emph{and the source. }This is, of course, just what one would expect
in a well-defined theory. If the source changes, we have a different
experiment, and would therefore expect a different solution. What
Earman-Norton call \textquotedbl{}radical local indeterminism\textquotedbl{}
is merely the correct action of a symmetry principle. General active
diffeomorphisms produce models of possible new experiments with new
sources and therefore new solutions. 

Thus the Earman-Newton treatment of indeterminism does not generalize
Einstein's version of the hole argument. Without an exact match of
their active \textquotedbl{}hole diffeomorphisms\textquotedbl{} to
an energy-momentum source, the Earman-Norton argument fails. With
that match, it simply replicates Einstein's version.

The Newtonian denial of Leibniz Equivalence avoids the unfortunate
consequences argued by Earman-Norton. Also, unlike a researcher who
adopts Leibniz Equivalence and therefore must deny substantivalism
and accept Leibnizian relativism, a researcher adopting Newton Equivalence
is not thereby committed to either side in the substantivalist-relativist
debate.

\section{Conclusion}

\label{sec:conclusion}Leibniz Equivalence and Newton Equivalence
are two different interpretations of the set of agreed upon facts
and practices listed in \prettyref{sec:summary}. 

Newton Equivalence interprets active diffeomorphisms in a manner consistent
with current theoretical and experimental physics. Active diffeomorphisms
are taken to be symmetry operations that change manifold objects and
therefore change the physical situation being modeled. But this changed
physical situation is one that is equally possible, that does not
violate the generally covariant laws of nature. Newton Equivalence
also permits a researcher to remain agnostic about the relativist-substantivalist
debate, thus avoiding restriction of the choices available to a theorist
facing current problems in theoretical physics such as a quantum theory
of gravity.

Leibniz Equivalence is based on an unjustified extrapolation of Leibnizian
relativism, applying it to all active diffeomorphisms, including localized
active diffeomorphisms that are non-identity only in limited regions,
and those that may distort the manifold rather than just translate
it. This extrapolation requires redefinition of the terms \textquotedbl{}physical
situation\textquotedbl{} and \textquotedbl{}observationally indistinguishable,\textquotedbl{}
giving them meanings inconsistent with the usual practice in differential
geometry and physics. Also, a researcher adopting Leibniz Equivalence
must deny substantivalism, thus restricting the choices available
to theorists.

These considerations would seem to favor Newton Equivalence. It is
a straightforward and unencumbered interpretation of active diffeomorphisms,
one that is consistent with the rules of differential geometry and
with current practice in theoretical and experimental physics. 

\bibliographystyle{spbasic}
\bibliography{ODJ}

\end{document}